\documentclass[pdflatex,sn-mathphys-ay]{sn-jnl}

\usepackage{graphicx}
\usepackage{amsmath,amssymb}
\usepackage{booktabs}
\usepackage{array}
\usepackage{tabularx}
\usepackage{multirow}
\usepackage{placeins}

\AtBeginDocument{%
  \hypersetup{colorlinks=true,citecolor=black,linkcolor=black,urlcolor=black}%
}

\graphicspath{{figures/}}
\begin{document}

\title[Centaur longevity revisited]{Centaur Longevity Revisited: Lifetime Dispersion and the Bailey \& Malhotra Classifications under Modern Orbital Solutions}

\author*[1]{\fnm{Naresh} \sur{Prasanna}}\email{Nareshprasanna06@gmail.com}
\author[2]{\fnm{Chrisphin} \sur{Karthick}}

\affil*[1]{\orgname{Knowledgeum Academy}, \orgaddress{\city{Bengaluru}, \postcode{560070}, \state{Karnataka}, \country{India}}}
\affil[2]{\orgname{Indian Institute of Astrophysics}, \orgaddress{\city{Bengaluru}, \postcode{560034}, \country{India}}}

\abstract{\unboldmath
Centaurs occupy dynamically unstable orbits between Jupiter and Neptune and are typically removed by giant-planet encounters within a few Myr. Bailey and Malhotra (hereafter BM09) classified the known Centaurs, from 2007 orbits, into a short-lived diffusing class (D), a resonance-hopping class (R), and a long-lived, quasi-stable class (Q); their sole Q object, 2005 TH173, was fit to a 17\,d arc.

We revisit both halves of that analysis on modern Jet Propulsion Laboratory Small-Body Database (SBDB) orbits (retrieved 2026-07-26), integrating clone ensembles of each object with the REBOUND WHFast $N$-body integrator. We ask (i) whether tighter orbital precision narrows the dispersion in clone lifetimes, and (ii) whether the BM09 labels survive on updated elements.

Among 65 genuine Centaurs with arc $\ge 30$\,d and no censored clones, $\sigma(\log_{10}\mathrm{lifetime})=0.54$\,dex (a factor of 3.5). Conditioned on present-day $a$, $e$, $i$, and $q$, this dispersion is insensitive to relative semimajor-axis uncertainty (slope $=-0.002$\,dex per decade in $\sigma_a/a$; 95\% confidence interval $-0.034$ to $+0.030$) --- a chaos floor.

On the D/Q axis, 49/50 comparable objects retain their BM09 class; 2005 TH173 is the exception (Q$\to$D): 80/100 clones escape before 10\,Myr (median 2.99\,Myr), and a 40\,Myr integration of the same initial conditions returns 88\,D / 12\,R / 0\,Q. None of the 30 multi-opposition 2007 orbits changes class; two short-arc objects leave the Centaur zone.}


\maketitle

\section*{Acknowledgements}
Jet Propulsion Laboratory (JPL) DE441 \citep{2021AJ....161..105P} and the Small-Body Database (SBDB); REBOUND \citep{2012A&A...537A.128R,2019MNRAS.485.5490R,2015MNRAS.452..376R}. We thank the authors of BM09 for a Table~2 extract that made an object-by-object reclassification possible.

\section{Introduction}
\label{sec:intro}

Centaurs are small icy bodies --- typically a few to a few hundred kilometres across --- that orbit the Sun in the region spanned by the four giant planets, rather than in the main asteroid belt or the distant Kuiper Belt. Two numbers fix the basic shape of an orbit: perihelion, the closest distance an object ever comes to the Sun, and semimajor axis, roughly its average distance from the Sun over one full orbit. A Centaur is defined by having a perihelion beyond Jupiter's orbit and a semimajor axis inside Neptune's, so its path threads between the giant planets rather than settling into a stable orbit the way a planet does. That position is precarious rather than permanent: repeated close encounters with Jupiter, Saturn, Uranus, and Neptune perturb a Centaur's orbit on timescales of only a few million years, eventually sending it either inward, where it can become a short-period comet, or back outward to the distant, icy region it likely came from. Centaurs are therefore studied both as a transient population linking two much larger reservoirs of small bodies, and as a natural laboratory for chaotic gravitational dynamics.

Centaurs occupy this dynamically crowded region between the giant planets, though survey papers vary somewhat in the adopted boundary \citep{2008ssbn.book...43G}. They form the dynamical bridge between the trans-Neptunian region --- the disk of icy bodies, including the Kuiper Belt, that lies beyond Neptune --- and the Jupiter-family comets, the short-period comets whose orbits are shaped mainly by Jupiter \citep{2015SSRv..197..191D,1997Sci...276.1670D,1997Icar..127...13L,2008ApJ...687..714V,2019ApJ...883L..25S}.

Their dynamical lifetimes are correspondingly short: encounters with the four giant planets typically remove a Centaur within a few Myr (millions of years) \citep{2007Icar..190..224D,2020CeMDA.132...36D,2004MNRAS.354..798H,2003AJ....126.3122T,2021Icar..35814201R}. Numerical integrations repeatedly recover two regimes: a random walk in $a$, in which the semimajor axis $a$ drifts gradually under the cumulative effect of many small gravitational kicks, and extended episodes of resonance sticking near mean-motion commensurabilities, in which an object's orbital period locks into a simple ratio with a planet's period (for example 2:1) and its semimajor axis holds nearly steady for a time before being kicked free again \citep{2007Icar..190..224D,2003AJ....126.3122T}.

BM09 \citep{2009Icar..203..155B} formalized this split using Hurst analysis --- a statistical technique, borrowed from time-series analysis, that distinguishes a smoothly diffusing signal from one with long memory or abrupt jumps --- applied to the $a(t)$ time series, the semimajor axis tracked over the course of an integration, together with a $\approx$22\,Myr long-life cut, defining
\begin{itemize}
\item \textbf{D} --- short-lived diffusion in $a(t)$;
\item \textbf{R} --- resonance hopping, admissible at any lifetime;
\item \textbf{Q} --- long-lived, nearly flat $a(t)$.
\end{itemize}
Table~2 of BM09 drew on Minor Planet Center (MPC) orbits at epoch 2007 March~6, several fit to arcs of only days. Their single Q Centaur, 2005 TH173, occupied a near-circular orbit between Saturn and Uranus on a 17\,d arc.

Such a label is only as reliable as the orbit it is drawn from. Subsequent studies mapped Centaur source regions, typical lifetimes, and end states without re-applying BM09's taxonomy object-by-object \citep{2005MNRAS.361.1345E,2007Icar..190..224D,2010A&A...519A.112D,2012MNRAS.420.3396B,2015SSRv..197..191D,2018AJ....155....2W,2019AJ....158..132N,2020CeMDA.132...36D}; mean lifetimes remain a few Myr and increase with inclination and perihelion \citep{2020CeMDA.132...36D,2004MNRAS.354..798H}. We do not repeat that population-level analysis; instead we pose two narrower questions against present-day orbits.

\begin{enumerate}
\item Does improved orbital precision reduce the dispersion in simulated lifetime among clones of a given Centaur?
\item Do the BM09 Table~2 classifications survive under modern JPL SBDB orbits at SBDB osculating epoch (2026-06-09 for most objects) and a stated REBOUND WHFast pipeline?
\end{enumerate}

A clone is a simulated realization of an object's orbit, offset from the published solution by an amount consistent with its formal uncertainty --- in effect, one plausible alternative orbit for the same object, drawn from the range the observations allow. Clone ensembles thus sample lifetime and class as functions of position within the published error volume: by simulating many such alternative orbits and following each one forward in time, we can ask whether a classification is a robust property of the object itself or an artifact of exactly which orbit happened to be published. This is a sensitivity analysis, not a reproduction of BM09's integrator, planetary model, escape treatment, or Hurst-based classifier; a label change can therefore be attributed to the orbit update, the numerical pipeline, the escape criteria, the classification rule, or some combination thereof.

These two questions draw on different samples and yield different answers. For lifetime dispersion, objects with arc $\ge 30$\,d already sit on a measured floor that is insensitive to $\sigma_a/a$, the fractional uncertainty in the semimajor axis. For classification, 49/50 objects that remain Centaurs and are comparable on the D/Q axis retain their BM09 class; the exception is 2005 TH173 (Q$\to$D). The nine BM09 R objects form a separate comparison, since our rule assigns R only after 22\,Myr of survival whereas BM09's R is defined by resonance-hopping morphology; those labels are not tested here. No D/Q label changes among the 2007 orbits already based on multiple oppositions.

Production tables use a lifetime-first rule (classifier v2); an earlier Hurst-gated rule (v1) serves as a robustness check in Appendix~\ref{app:v1}.

\paragraph{Terminology.}
(i)~Dex: a $\log_{10}$ unit; a difference of $d$ dex corresponds to a factor of $10^{d}$ in lifetime (0.54\,dex $\approx \times 3.5$).
(ii)~Right-censored clone: a clone still bound when the integration halted, so its true lifetime exceeds the recorded value.
(iii)~Diagonal clones: independent $1\sigma$ perturbations of each orbital element, neglecting cross-covariances.
(iv)~30\,d arc cut: a data-quality threshold, not a dynamical boundary.
(v)~10\,Myr integration: can confirm an early escape but does not test the 22\,Myr Q threshold.

\section{Methods}

For each object we draw a set of plausible starting orbits, integrate them under a fixed $N$-body model, record each clone's lifetime, and classify accordingly. Within-object lifetime dispersion addresses question~(1); the modal clone class, together with the full D/R/Q counts, addresses question~(2).

\subsection{Sample}
\label{sec:sample}

Osculating elements --- the instantaneous Keplerian orbit an object would follow if all perturbations vanished at that instant --- at SBDB osculating epoch (2026-06-09 for most objects) are drawn from the JPL SBDB API (retrieved 2026-07-26); BM09 Table~2 names are matched by provisional designation. 2005 TH173 remains unnumbered (18 observations, 366\,d arc, condition code $U=6$ --- the Minor Planet Center's orbit-uncertainty flag, on which a higher value marks a less well-determined orbit). 1995 SN55 is now numbered as the trans-Neptunian object (523731)~2014~OK394. 2000 SN331 (BM09 class R; 1\,d arc) has no SBDB match and is omitted; the Neptune Trojan 2006 RJ103 is excluded from the dynamics sample, following BM09.

Table~\ref{tab:sample} traces the reduction from catalogue to analysis samples. The 30\,d cut is a data-quality gate --- below it the published solution is often a single-apparition fit, and BM09's own Q object entered on a 17\,d arc --- not evidence that every object above threshold has converged to a stable solution. Within the complete-case sample (\S\ref{sec:dispersion}), $\log_{10}(\mathrm{arc})$ and median clone lifetime are uncorrelated ($\rho=-0.14$, $p=0.26$, $N=65$), consistent with treating the cut as a sample restriction rather than a dynamical criterion.

\begin{table}[t]
\centering
\small
\caption{Sample selection. The 219-row dispersion ingest is $163+61-5$: nine of the 172 Appendix~\ref{app:f} survey objects lack a usable $\sigma_a$ and are unused in the regression.\label{tab:sample}}
\begin{tabular}{@{}lr@{}}
\toprule
Stage & $N$ objects \\
\midrule
SBDB Centaur flag (query 2026-08-14) & 1046 \\
Usable covariance & 970 \\
Genuine Centaur ($q>5.2$\,AU, $a<30.07$\,AU) & 410 \\
Arc $\ge 30$\,d (catalogue projection) & 355 \\
Stratified survey (Appendix~\ref{app:f}) & 172 \\
Survey objects with usable $\sigma_a$ & 163 \\
BM09 Table~2 with modern SBDB & 61 \\
Overlap of survey and BM09 & 5 \\
Pooled clone ensembles ($163+61-5$) & 219 \\
Fail membership ($q\le 5.2$\,AU or $a\ge 30.07$\,AU) & 101 \\
Genuine Centaurs in the pooled ensembles & 118 \\
Complete-case lifetime dispersion (no censored clones, arc $\ge 30$\,d) & 65 \\
Kaplan--Meier lifetime dispersion (arc $\ge 30$\,d) & 96 \\
Remain Centaurs and comparable on D/Q & 50 \\
\bottomrule
\end{tabular}
\end{table}

Table~\ref{tab:bm09diff} lists what differs from BM09. Any BM09$\to$modern label change is the joint product of these choices.

\begin{table}[t]
\centering
\small
\caption{Differences from BM09.\label{tab:bm09diff}}
\begin{tabular}{@{}p{0.26\columnwidth}p{0.66\columnwidth}@{}}
\toprule
Component & Difference and likely effect \\
\midrule
Orbital elements & 2026 SBDB vs.\ 2007 MPC; can move objects between dynamical niches (TH173, SN55). \\
Integrator & REBOUND WHFast vs.\ BM09's setup; affects close-encounter timing and lifetimes. \\
Planets & Four giants, DE441 start, then freely evolved; no continuous ephemeris forcing and no terrestrial planets. \\
Escape cuts & Stated heliocentric-distance, perihelion, and Hill-sphere cuts; change loss times. \\
Classifier & Lifetime-first tree vs.\ BM09 Hurst-gated rules; affects long-lived R/Q. \\
Clone ensembles & Independent one-sigma draws, omitting correlations. \\
$T_{\max}$ & 10\,Myr for the TH173 early-escape test; 40\,Myr to resolve R versus Q. \\
\bottomrule
\end{tabular}
\end{table}

\subsection{Orbit generation}
\label{sec:clones}

Where element uncertainties are available, we draw independent Gaussian perturbations on the diagonal $1\sigma$ errors, omitting SBDB cross-covariances. The resulting ensemble --- hereafter diagonal clones --- is a reproducible sensitivity baseline rather than a draw from the full observational posterior.

For the Uranus-crossing TH173 and the still short-arc 2003 QP112 ($\sigma_a\approx 5.8$\,AU, $\sigma_e\approx 0.31$ at its own osculating epoch 2003-09-24), neglecting correlations can place clones in dynamically improbable corners of $(a,e,i)$-space, primarily affecting minority R/Q fractions. This is less consequential for the TH173 Q test: 80/100 production clones escape before 10\,Myr and the nominal clone is itself short-lived. A 20-clone TH173 ensemble drawn from the full SBDB covariance ($T_{\max}=10$\,Myr, seed 87) is reported in \S\ref{sec:th173} as a Q-rejection check only. Clone index 0 is always the published nominal; seed 87 is used throughout.

Clone counts are matched to the precision required: with $N=20$, a single clone shifts a reported fraction by 5\%; with $N=5$, by 20\%. The modal class can appear stable while minority fractions remain poorly resolved, so we report full D/R/Q counts alongside the mode without quoting bootstrap intervals on those fractions.

\subsection{Integration}
\label{sec:integrator}

The Sun and four giant planets are integrated with REBOUND's WHFast symplectic integrator \citep{2012A&A...537A.128R,2015MNRAS.452..376R} --- a Wisdom--Holman mixed-variable map \citep{1991AJ....102.1528W} --- at timestep $dt=0.25$\,yr. Centaur lifetimes of the kind classified by BM09 are governed by giant-planet scattering on Myr timescales, so resolving terrestrial close encounters is unnecessary for the production runs. Giant-planet heliocentric states are initialized from JPL DE441 \citep{2021AJ....161..105P}, a numerically integrated planetary and lunar ephemeris, at epoch J2000 (JD 2451545.0), rotated from the International Celestial Reference Frame (ICRF) --- the standard reference grid for sky coordinates --- to the J2000 ecliptic, and thereafter evolved freely with the massless Centaur under no continuous ephemeris forcing --- i.e., the four-planet $N$-body problem, as in the classical Centaur experiments \citep{2009Icar..203..155B,2007Icar..190..224D,2004MNRAS.354..798H,2003AJ....126.3122T}. The Centaur's osculating elements are those of the SBDB osculating epoch (2026-06-09, JD 2461200.5, for most objects); both the planets and the Centaur are inserted at simulation $t=0$, so their source epochs differ by $\approx 26.4$\,yr for most objects. For 2003 QP112 the osculating epoch is 2003-09-24 (JD 2452906.5) and the offset from J2000 is $\approx 3.7$\,yr.

Particles are removed upon reaching heliocentric distance $r>10^{4}$\,AU, perihelion $q<2.5$\,AU, or a giant-planet Hill sphere --- the region around a planet within which its own gravity dominates over the Sun's, so a clone entering it counts as a close encounter --- as the baseline criterion set. The $q<2.5$\,AU criterion is a practical proxy for Jupiter-family-comet entry / inner-system removal \citep{2018P&SS..158....6F,2004MNRAS.354..798H,1997Icar..127...13L}, not a physical disruption model; runs omitting the Hill-sphere cut are sensitivity checks only.

\subsection{Classification}
\label{sec:classifier}

Following BM09's cadence, heliocentric $a$ is recorded every 300\,yr; windowed scatter in $a(t)$ yields a Hurst exponent $H$ and a linear $R^{2}$, retained here as diagnostics only. This cadence is coarse relative to encounter timescales --- BM09's classes describe the long-window $a(t)$ series, not individual Hill-sphere passages \citep{1994Icar..108...18L}.

A clone's event time is the elapsed time to its first escape criterion ($r>10^{4}$\,AU, $q<2.5$\,AU, or a giant-planet Hill sphere), not the time spent inside the Centaur zone ($q>5.2$\,AU, $a<30.07$\,AU). A clone still bound when integration halts is right-censored; a confirmed early escape satisfies an escape criterion before $T_{\max}$. Survival to a 10\,Myr stop does not test the 22\,Myr Q threshold --- such survivors remain unclassified on R versus Q pending a longer integration. The long-lived QP112 clone of Fig.~\ref{fig:at} first reaches $a\ge 30.07$\,AU well before 22\,Myr, yet remains bound to 40\,Myr under the production escape cuts.

Table~\ref{tab:classes} compares BM09's categories with the production rule used here. The production rule (v2) is lifetime-first (Table~\ref{tab:classes}).

\begin{table}[t]
\centering
\small
\caption{Classification rules. Direct BM09 comparison is valid on the D/Q axis only.\label{tab:classes}}
\begin{tabular}{@{}p{0.10\columnwidth}p{0.40\columnwidth}p{0.40\columnwidth}@{}}
\toprule
Class & BM09 & This work (v2) \\
\midrule
D & Short-lived diffusion of $a(t)$ & Lifetime $<22$\,Myr \\
R & Resonance hopping, any lifetime & Lifetime $\ge 22$\,Myr and not flat \\
Q & Long-lived, nearly flat $a(t)$ & Lifetime $\ge 22$\,Myr and flat \\
\bottomrule
\end{tabular}
\end{table}

Flatness is defined as $\sigma_a/\langle a\rangle<0.03$ on the stored $a(t)$ series. Ties (equal D and R counts) are reported explicitly and, for table-sorting purposes only, list D first; a 10/10 split is not treated as a D majority.

BM09 associated short lifetimes with diffusion (D), long lifetimes with resonance hopping (R), and long-lived flat $a(t)$ with Q. Given population mean lifetimes of a few Myr \citep{2007Icar..190..224D,2020CeMDA.132...36D,2004MNRAS.354..798H}, 22\,Myr lies in the high tail; we adopt this threshold so that D and Q remain comparable with BM09's Table~2. Our R differs fundamentally from BM09's: theirs is a morphological designation assignable at any lifetime, whereas ours requires $\ge 22$\,Myr survival. Direct comparison is therefore valid for D and Q only.

An earlier Hurst-gated rule (v1) assigned D whenever the linear Hurst fit was good ($R^{2}\ge 0.85$), irrespective of survival past 22\,Myr. Production tables use v2; the v1 comparison appears in Appendix~\ref{app:v1}.

\subsection{Statistical analysis}
\label{sec:dispersion}

To separate observational from intrinsic uncertainty we pool two clone campaigns: 172 objects from a stratified SBDB Centaur survey (10 clones per object, $T_{\max}=25$\,Myr, with 40\,Myr extensions for censored clones), and the 61 BM09 Table~2 objects under the same protocol. Five names appear in both campaigns. Nine survey objects lack a usable SBDB semimajor-axis uncertainty (all have $a$ but $\sigma_a$ missing) and drop out at ingest, leaving 219 unique ensembles ($163+61-5$; Table~\ref{tab:sample}). Where an object appears in both campaigns, the longer integration supersedes the shorter.

The outcome variable is within-object dispersion in $\log_{10}$ lifetime. For ensembles with no right-censored survivors we compute the sample standard deviation of $\log_{10}$ lifetime directly; for ensembles with censoring --- where some clones never escaped before the integration stopped --- we instead build a Kaplan--Meier curve, the standard survival-analysis estimator for a distribution with such censored observations, take its interquartile range in $\log_{10}$ lifetime, and divide by 1.349, the constant that converts a normal distribution's interquartile range into an equivalent standard deviation, to place it on a comparable scale. Objects whose clones were removed at the first output step ($t=0$) are excluded, as such zeros reflect the escape criterion rather than dynamical dispersion.

The measured leftover dispersion is the empirical quantity; the chaos-floor interpretation is the proposed explanation. Giant-planet-crossing Centaurs are strongly chaotic \citep{2009Icar..203..155B}. The 39-object proxy in \S\ref{sec:lyapunovresults} spans $\approx 6.4\times 10^{2}$--$1.4\times 10^{5}$\,yr, far shorter than our 25--40\,Myr integration windows. Once an integration extends well past that timescale, nearby clones lose memory of their initial offset and sample the same chaotic region; a flat relation between dispersion and $\sigma_a/a$ is a population-scale measurement of that saturation, not a novel detection of chaos per se.

Orbit quality is quantified as the relative semimajor-axis uncertainty $\sigma_a/a$ from the SBDB covariance. The primary regression is dispersion against $\log_{10}(\sigma_a/a)$, conditioned on osculating $(a,e,i,q)$, since these elements independently predict dispersion and are entangled with observability. ``Per decade'' denotes per tenfold change in $\sigma_a/a$, not a ten-year interval. Inference proceeds from the slope excluded at 95\% confidence; a null or weak association bounds, but does not prove, exact independence.

\subsection{Lyapunov-time proxy}
\label{sec:lyapunovproxy}

To relate the leftover dispersion to a standard chaotic timescale without variational Lyapunov integrations for every object, we ran a stratified pilot on 39 objects from the dispersion sample (10 diagonal clones, seed 87, $T_{\max}=25$\,Myr), recording osculating $a$ at snapshots $t\in\{0.002,0.005,0.01,0.05,0.1,0.25,0.5,1,2,5,10\}$\,Myr while clones remained bound. The ensemble relative dispersion $\mathrm{std}(a)/\langle a\rangle$ grows approximately exponentially over the first 0.1\,Myr; we estimate a proxy timescale $\tau_{\mathrm{proxy}}$ as the geometric mean of pairwise growth rates within that window, anchored to the initial $\sigma_a/a$ from the SBDB covariance.

$\tau_{\mathrm{proxy}}$ is not a finite-time Lyapunov exponent obtained from variational equations or from MEGNO (the Mean Exponential Growth factor of Nearby Orbits, a standard numerical chaos indicator); it measures how rapidly a finite diagonal clone ensemble, seeded at observational uncertainty, diverges in osculating $a$.

\subsection{Campaigns and encounter check}
\label{sec:campaigns}

Campaign length is matched to the question at hand (Table~\ref{tab:exps}). TH173 production uses $T_{\max}=10$\,Myr as an early-escape test: a clone meeting an escape criterion before 10\,Myr cannot subsequently satisfy the 22\,Myr Q threshold. A clone still bound at 10\,Myr is right-censored and remains unclassified (neither ruling out nor confirming later survival to 22\,Myr). Resolving R versus Q requires $T_{\max}\ge 22$\,Myr, so the Q/R matrix and census extensions adopt 40\,Myr.

Since the modern TH173 orbit is Uranus-crossing, plain WHFast performance near close approaches is not guaranteed \citep{1999MNRAS.304..793C}; we therefore re-integrated the first 20 production clones (seed 87) with REBOUND Mercurius \citep{2019MNRAS.485.5490R} at $T_{\max}=10$\,Myr. Both integrators return zero Q clones on this subset, though individual clone trajectories do not carry over between the two, and median lifetimes differ (WHFast 3.86\,Myr; Mercurius 5.56\,Myr; cf.\ 2.99\,Myr for the full 100-clone WHFast ensemble). This comparison tests the ensemble-level Q verdict, not agreement of any single chaotic trajectory.

\begin{table*}[t]
\centering
\small
\caption{Simulation campaigns. $N$ is clones per object except where noted.\label{tab:exps}}
\begin{tabularx}{\textwidth}{@{}>{\raggedright\arraybackslash}p{3.15cm}>{\raggedright\arraybackslash}p{2.55cm}cc>{\raggedright\arraybackslash}X@{}}
\toprule
Campaign & Sample & $N$ & $T_{\max}$ & Purpose \\
\midrule
TH173 10\,Myr (primary) & TH173 & 100 & 10\,Myr & Early-escape Q test; 20 clones remain censored \\
TH173 40\,Myr (primary) & TH173 & 100 & 40\,Myr & Resolve R versus Q on the same seed \\
Q/R matrix (primary) & BM09 Q$+$R (10) & 20 & 40\,Myr & Clone counts and modal class \\
BM09 Table~2 census (primary) & All BM09 (61) & 10 & 25\,Myr & Modal class for the published sample \\
Census extensions & Censored clones & --- & 40\,Myr & Resolve R versus Q for survivors \\
Covariance Q-test & TH173 & 20 & 10\,Myr & Full SBDB covariance; Q rejection only \\
Orbit-update control & Same 10 Q/R & 1/branch & 40\,Myr & Orbit revision versus pipeline \\
Dispersion analysis & Survey$+$BM09 (219) & 10 & 25--40\,Myr & Lifetime dispersion vs.\ $\sigma_a/a$ \\
Lyapunov proxy & Pilot (39) & 10 & 25\,Myr & Early $a$-dispersion vs.\ leftover dispersion \\
\bottomrule
\end{tabularx}
\end{table*}

\subsection{Software and AI use in the research}
\label{sec:software}

Integrations, clone generation, the classification rule, campaign scheduling, statistical aggregation, and figures are implemented in Python using REBOUND \citep{2012A&A...537A.128R,2019MNRAS.485.5490R,2015MNRAS.452..376R}; source code and archived production JSON accompany this article.

The first author developed this code in Cursor (Anysphere), using Composer~2.5 and Grok~4.6 to draft and refactor the REBOUND driver, clone generator, classifier, campaign scheduler, statistical aggregation, and figure scripts. The scientific specification --- integrator and timestep, escape criteria, the 22\,Myr threshold, Centaur membership boundaries, clone counts, random seed 87, and the regression model --- was fixed by the authors and is not a tool output. The models did not generate, alter, or impute observational or integration data, nor invent numerical results; every reported number, table, and figure derives from archived campaign JSON via the released scripts. Neither tool is an author of this work.

Three code checks are reported. A known-answer battery applies the production rule to six synthetic $a(t)$ series with predetermined labels, recovering the intended label in all six cases. A second audit tabulates raw per-clone labels across every archived run (rather than modal classes only), which exposed the v1 defect discussed in Appendix~\ref{app:v1}. A third, independent pair of scripts regenerates all tabulated quantities from per-object JSON.

\section{Results}

\subsection{Lifetime dispersion and orbit precision}
\label{sec:chaosfloor}

Among genuine Centaurs with arc $\ge 30$\,d, clone-lifetime dispersion tracks the object's dynamical state rather than the precision with which that state is measured (Table~\ref{tab:stats}).

Of the 219 pooled ensembles, 101 fail the Centaur membership test ($q>5.2$\,AU, $a<30.07$\,AU); among the remaining 118 genuine Centaurs, 52 ensembles retain at least one clone alive at the integration cap.

The complete-case sample comprises 65 Centaurs with no censored clones. $\log_{10}(\sigma_a/a)$ spans $-7.5$ to $-0.5$, and observation arcs span 54--47\,770\,d. Mean dispersion is 0.54\,dex (95\% confidence interval (CI) 0.50--0.58; median 0.51) --- a factor of 3.5 in lifetime --- persisting even as $\sigma_a/a$ approaches one part in $3\times 10^{7}$. Conditioned on osculating $(a,e,i,q)$, the coefficient on $\log_{10}(\sigma_a/a)$ is $-0.002$\,dex per decade (95\% CI $-0.034$ to $+0.030$; $p=0.91$); $a$ and $i$ independently predict dispersion ($p=0.045$ and $p=0.028$ respectively), while $\sigma_a/a$ does not. A partial Spearman correlation between dispersion and $\sigma_a/a$, conditioned on the same osculating elements, gives $\rho=-0.055$ ($p=0.66$; Table~\ref{tab:stats}), consistent with the regression above. Across the observed range, this bound permits better astrometry to alter lifetime dispersion by no more than a factor of $\approx 1.7$ (point estimate $\approx 1.0$) --- a bound, not a demonstration of exact independence.

\begin{figure*}[tbp]
\centering
\includegraphics[width=0.96\textwidth]{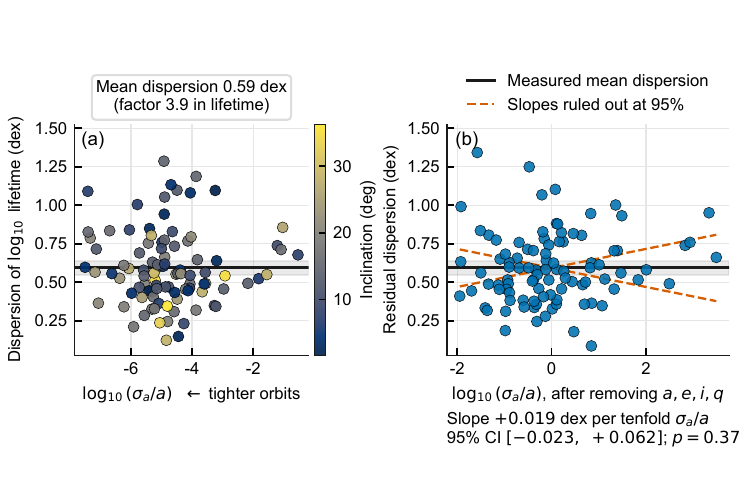}
\caption{Within-object dispersion in $\log_{10}$ lifetime versus orbit quality for 96 Centaurs with arc $\ge 30$\,d (Kaplan--Meier estimator). (a)~Observed relation, coloured by inclination; the horizontal band marks the measured leftover dispersion (0.59\,dex, $\approx\times 3.9$ in lifetime). (b)~Residuals after conditioning on osculating $(a,e,i,q)$; dashed lines bound the slopes excluded at 95\% confidence. $\sigma_a/a$ spans seven orders of magnitude without a corresponding rise in dispersion.\label{fig:chaosfloor}}
\end{figure*}

A Kaplan--Meier estimator over 96 objects gives mean dispersion 0.59\,dex (median 0.56) and a coefficient of $+0.019$\,dex per decade in $\sigma_a/a$ (95\% CI $-0.023$ to $+0.062$; $p=0.37$; Fig.~\ref{fig:chaosfloor}).

Of 410 genuine Centaurs, 355 already have arc $\ge 30$\,d; the remaining 55 lack a robust first solution. Continued astrometry can supply that solution, but --- under the derived bound --- cannot further reduce lifetime dispersion among the 355 already past the threshold.

\begin{table*}[t]
\centering
\small
\caption{Lifetime-dispersion regressions. The slope is per tenfold change in $\sigma_a/a$, after accounting for osculating $(a,e,i,q)$ except in the unadjusted complete-case mean.\label{tab:stats}}
\begin{tabularx}{\textwidth}{@{}>{\raggedright\arraybackslash}Xrrll@{}}
\toprule
Analysis & Objects & Clones/obj. & Estimate & 95\% CI / $p$ \\
\midrule
Complete-case mean dispersion & 65 & 10 & 0.54\,dex & 0.50--0.58 \\
Complete-case slope on $\log_{10}(\sigma_a/a)$ & 65 & 10 & $-0.002$\,dex/decade & $-0.034$ to $+0.030$; $p=0.91$ \\
Kaplan--Meier mean dispersion & 96 & 10 & 0.59\,dex & median 0.56 \\
Kaplan--Meier slope on $\log_{10}(\sigma_a/a)$ & 96 & 10 & $+0.019$\,dex/decade & $-0.023$ to $+0.062$; $p=0.37$ \\
Partial Spearman vs $\sigma_a/a$ & 65 & 10 & $\rho=-0.055$ & $p=0.66$ \\
\bottomrule
\end{tabularx}
\end{table*}

\subsection{Ensemble divergence timescale}
\label{sec:lyapunovresults}

\begin{figure}[tbp]
\centering
\includegraphics[width=\columnwidth]{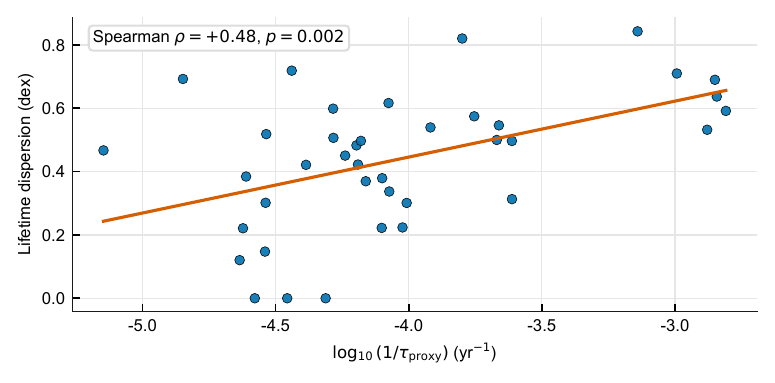}
\caption{Clone-ensemble chaos proxy (10 diagonal clones per object): lifetime dispersion versus $\log_{10}(1/\tau_{\mathrm{proxy}})$ for the 39-object pilot ($\rho=+0.48$, $p=0.002$); faster ensemble divergence predicts larger leftover dispersion.\label{fig:lyapunovproxy}}
\end{figure}

Across the 39-object pilot, $\tau_{\mathrm{proxy}}$ spans $\approx 6.4\times 10^{2}$--$1.4\times 10^{5}$\,yr. Faster local divergence correlates with larger lifetime dispersion: Spearman($1/\tau_{\mathrm{proxy}}$, dispersion) $\rho=+0.48$ ($p=0.002$, $N=39$; Fig.~\ref{fig:lyapunovproxy}).

\subsection{The BM09 Q object: 2005 TH173}
\label{sec:th173}

BM09's sole Q Centaur furnishes the clearest single-object revision. TH173 entered BM09 Table~2 on a 17\,d arc, well short of the regime where the chaos floor applies. The argument that follows rests on BM09's own definition of Q --- lifetime $\ge 22$\,Myr --- independent of classifier v2's particulars. Two conditions must hold for the Q label to fail, and both do: the modern solution is dynamically distinct from the one BM09 classified, and clones drawn from it fail to survive long enough to qualify.

\begin{figure}[tbp]
\centering
\includegraphics[width=\columnwidth]{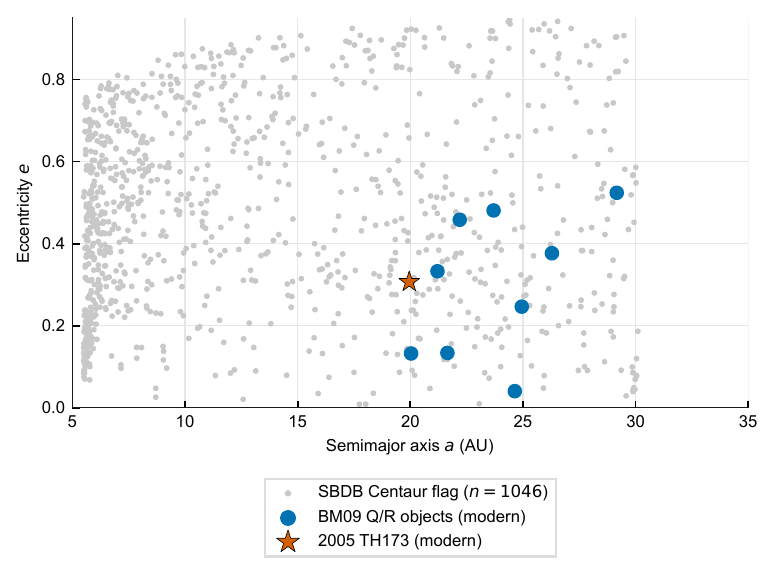}
\caption{Modern osculating $(a,e)$ for 2005 TH173 (star) and the other nine BM09 Q/R objects with SBDB matches (filled circles), over all JPL SBDB objects flagged as Centaurs ($N=1046$, a catalogue class rather than a debiased census). Modern TH173 lies among the Uranus-crossing Centaurs, not in BM09's low-$e$ Saturn--Uranus niche.\label{fig:aectx}}
\end{figure}

Modern SBDB elements displace TH173 out of the low-eccentricity Saturn--Uranus niche (Table~\ref{tab:th173od}): $\Delta a=+4.228$\,AU and $\Delta e=+0.293$ rank among the largest revisions in the joined BM09 Table~2 sample. Figure~\ref{fig:aectx} situates the modern TH173 solution among Uranus-crossing Centaurs at $e\approx 0.31$.

\begin{table}[tbp]
\centering
\small
\caption{2005 TH173 orbital elements: BM09 (2007 MPC extract) versus JPL SBDB elements at SBDB osculating epoch 2026-06-09 (retrieved 2026-07-26). The last row is the classification outcome of this work (100 clones, 10\,Myr), not an SBDB field.\label{tab:th173od}}
\begin{tabular}{@{}lcc@{}}
\toprule
Quantity & BM09 & Modern \\
\midrule
$a$ (AU) & 15.724 & 19.952 \\
$e$ & 0.014 & 0.307 \\
$i$ (deg) & 15.7 & 13.48 \\
$q$ (AU) & 15.5 & 13.83 \\
Arc & 17\,d & 366\,d \\
Condition code & --- & $U=6$ \\
BM09 / this work & Q & Q rejected (80 escaped / 20 censored) \\
\bottomrule
\end{tabular}
\end{table}

Primary production (seed 87, 100 diagonal clones, WHFast, $T_{\max}=10$\,Myr): 80 clones escape before 10\,Myr, 20 remain bound and are right-censored; median lifetime 2.99\,Myr, mean 4.38\,Myr. The 80 early escapes are necessarily non-Q; the 20 survivors remain unclassified on R versus Q at this stop.

\begin{figure*}[tbp]
\centering
\includegraphics[width=0.98\textwidth]{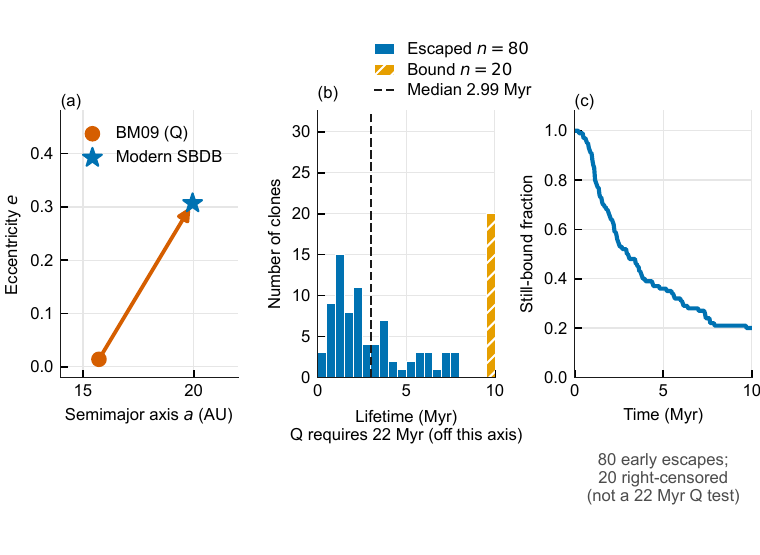}
\caption{2005 TH173 production ensemble. (a)~BM09 Table~2 $(a,e)$ versus modern SBDB $(a,e)$. (b)~Lifetime histogram, 100 diagonal clones, seed 87, WHFast, $T_{\max}=10$\,Myr; orange bars mark the 20 right-censored clones (the 22\,Myr Q cut lies off-scale). (c)~Ensemble survival fraction versus time --- survival at 10\,Myr is not itself a D, R, or Q label.\label{fig:th173}}
\end{figure*}

A second integration, regenerated from the same seed-87 initial conditions to $T_{\max}=40$\,Myr (not a bit-for-bit continuation of the 10\,Myr trajectories), tests the 22\,Myr threshold directly: 88\,D / 12\,R / 0\,Q (median 3.40\,Myr; fraction with lifetime $\ge 22$\,Myr $=0.12$). The longer window admits a minority surviving past 22\,Myr (class R) but still yields zero Q clones. Figure~\ref{fig:th173} shows the $(a,e)$ shift, the 10\,Myr lifetime histogram, and the survival curve. The primary median cited in the Abstract and Conclusions (2.99\,Myr) is from the 100-clone, 10\,Myr WHFast ensemble.

A 20-clone full-covariance test at 10\,Myr again returns 0\,Q (15 early escapes, 5 censored, median 3.82\,Myr): replacing diagonal $1\sigma$ draws with the joint covariance does not restore a Q verdict within this window.

\subsection{Do the published classifications survive?}

\begin{figure}[tbp]
\centering
\includegraphics[width=\columnwidth]{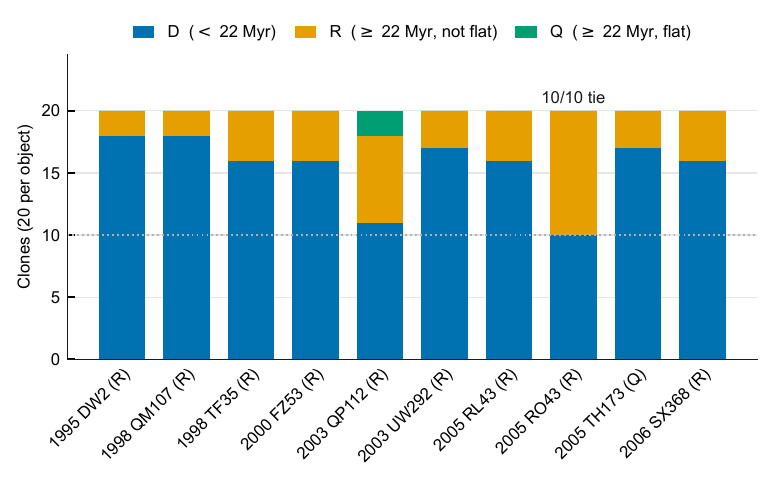}
\caption{Stacked clone counts (D/R/Q) for the ten BM09 Q/R objects with modern SBDB matches (20 clones per object, $T_{\max}=40$\,Myr). Nine objects have D as the modal class; RO43 is a 10\,D / 10\,R tie, shown as equal stacks rather than a D majority.\label{fig:qr}}
\end{figure}

Under the production rule (20 clones per object, $T_{\max}=40$\,Myr, diagonal SBDB clones, seed 87), nine of the ten BM09 Q/R objects with modern SBDB matches have D as the modal clone class (Table~\ref{tab:qr}; Fig.~\ref{fig:qr}); 2005 RO43 is a tie (10\,D / 10\,R; median 25.05\,Myr). Aggregated across all 200 clones: D $=155$, R $=43$, Q $=2$. Every object retains at least one R clone; only QP112 produces Q clones (2/20).

\begin{figure}[tbp]
\centering
\includegraphics[width=\columnwidth]{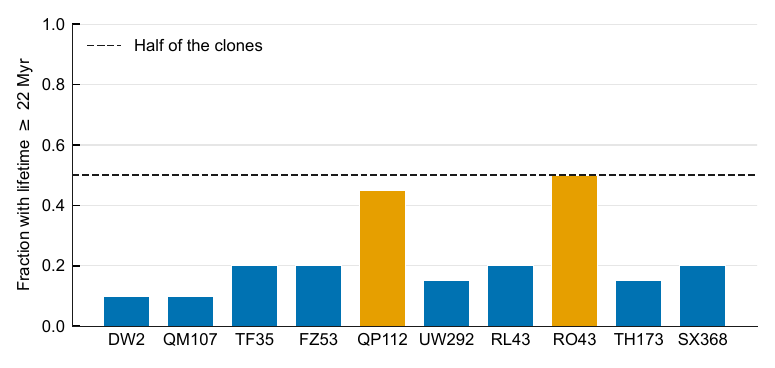}
\caption{Fraction of clones with lifetime $\ge 22$\,Myr for the ten ensembles of Fig.~\ref{fig:qr} and Table~\ref{tab:qr} (RO43 $=0.50$, tie; QP112 $=0.45$; remaining objects 0.10--0.20); the dashed line marks 0.5.\label{fig:longfrac}}
\end{figure}

QP112 carries a substantial long-lived minority (fraction $\ge 22$\,Myr $=0.45$; 7\,R, 2\,Q of 20 clones); Fig.~\ref{fig:longfrac} gives this fraction for all ten objects (RO43 $=0.50$; QP112 $=0.45$; remaining eight 0.10--0.20).

2000 FZ53 was the longest-lived object in the 32-Centaur sample of \citet{2004MNRAS.354..798H} (half-life 32\,Myr), yet on 2026 SBDB elements its modal class is D (16\,D / 4\,R) --- a long half-life under earlier elements does not guarantee a modal R class under modern ones.

\begin{figure*}[tbp]
\centering
\includegraphics[width=0.82\textwidth]{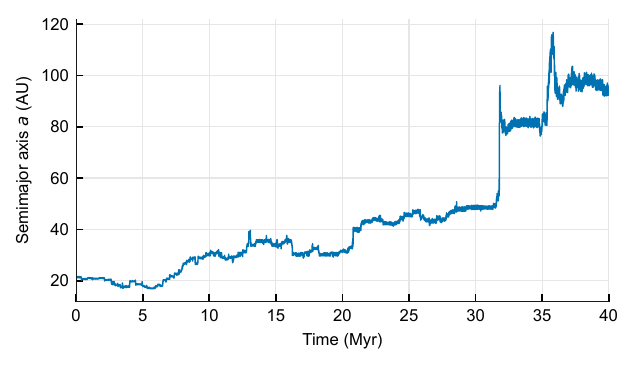}
\caption{Heliocentric $a(t)$ for 2003 QP112 clone 0 over a 40\,Myr WHFast integration (seed 87), regenerated from the archived production initial condition; this trajectory survives 40\,Myr and is class R under the production rule, though the QP112 ensemble as a whole is modal D (11\,D / 7\,R / 2\,Q).\label{fig:at}}
\end{figure*}

Figure~\ref{fig:at} regenerates $a(t)$ for one long-lived QP112 clone (index 0, seed 87); this regeneration survives 40\,Myr and is class R under the production rule, illustrating the plateau-and-jump morphology BM09 associated with resonance sticking \citep{2009Icar..203..155B,2007Icar..190..224D,2003AJ....126.3122T}. The QP112 production count (11\,D / 7\,R / 2\,Q; Table~\ref{tab:qr}) remains as reported.

\begin{table*}[tbp]
\centering
\small
\caption{BM09 Q/R sample under the production rule: 20 diagonal clones per object, 40\,Myr, seed 87. Modal class is a tie when D and R counts are equal. Aggregate: D$=$155, R$=$43, Q$=$2.\label{tab:qr}}
\begin{tabular}{lccrrr}
\toprule
Desig. & BM09 & Modal & D/R/Q & frac$\ge$22\,Myr & Med.\ life (Myr) \\
\midrule
1995 DW2 & R & D & 18/2/0 & 0.10 & 1.99 \\
1998 QM107 & R & D & 18/2/0 & 0.10 & 1.57 \\
1998 TF35 & R & D & 16/4/0 & 0.20 & 5.91 \\
2000 FZ53 & R & D & 16/4/0 & 0.20 & 4.87 \\
2003 QP112 & R & D & 11/7/2 & 0.45 & 15.24 \\
2003 UW292 & R & D & 17/3/0 & 0.15 & 6.18 \\
2005 RL43 & R & D & 16/4/0 & 0.20 & 9.61 \\
2005 RO43 & R & tie & 10/10/0 & 0.50 & 25.05 \\
2005 TH173 & Q & D & 17/3/0 & 0.15 & 2.54 \\
2006 SX368 & R & D & 16/4/0 & 0.20 & 8.35 \\
\bottomrule
\end{tabular}
\end{table*}

\subsection{Full BM09 Table~2 census}
\label{sec:census}

Of the 63 names in BM09 Table~2, one is the Neptune Trojan BM09 excluded from their own dynamics sample, and one (2000 SN331) lacks a modern SBDB solution. The remaining 61 each receive ten diagonal clones to $T_{\max}=25$\,Myr (seed 87); right-censored clones are re-integrated to 40\,Myr, which can only move a clone between R and Q, since survival to 25\,Myr already exceeds the 22\,Myr cut.

Two checks precede reading the census as a reclassification. The first is population membership: applying the Centaur zone ($q>5.2$\,AU, $a<30.07$\,AU) to both element sets, all 61 objects qualify under BM09's 2007 elements but only 59 do under modern ones (Table~\ref{tab:membership}). 1995 SN55 --- a Saturn-crossing Centaur on a 36\,d 2007 arc --- now sits at $a=42.6$\,AU, $q=35.5$\,AU on a 7708\,d arc, a detached trans-Neptunian object; 2002 FY36 crosses the outer boundary less dramatically. Both entered BM09 Table~2 on arcs of tens of days; none of the 36 multi-opposition objects leaves the population. Thirty of those 36 remain D/Q-comparable once the nine BM09 R objects are set aside; the 0/30 class-stability result below is that subset.

\begin{table}[t]
\centering
\small
\caption{BM09 Table~2 objects that leave the Centaur zone ($q>5.2$\,AU, $a<30.07$\,AU) under modern elements.\label{tab:membership}}
\begin{tabular}{@{}lrrrrl@{}}
\toprule
Desig. & $a$ (2007) & $q$ (2007) & $a$ (modern) & $q$ (modern) & Arc 2007 $\to$ now \\
\midrule
1995 SN55 & 23.56 & 7.94 & 42.61 & 35.53 & 36\,d $\to$ 7708\,d \\
2002 FY36 & 28.97 & 25.65 & 31.92 & 22.50 & 51\,d $\to$ 7774\,d \\
\bottomrule
\end{tabular}
\end{table}

The second check is class comparability: BM09's R denotes resonance hopping irrespective of lifetime, whereas our rule assigns R only above 22\,Myr survival. The shared label measures different quantities, so the R objects are excluded from the like-for-like comparison.

\begin{table}[t]
\centering
\small
\caption{BM09 Table~2 census under classifier v2: 61 objects, 10 diagonal clones each, $T_{\max}=25$\,Myr, seed 87. After excluding the two objects that leave the Centaur zone, 49 of 50 remaining D/Q-comparable objects keep their class; the sole change is TH173. The R row is not a like-for-like comparison.\label{tab:census}}
\begin{tabular}{@{}lccccc@{}}
\toprule
BM09 class & $n$ & D & R & Q & Retained \\
\midrule
D & 51 & 50 & 0 & 1 & 98\% \\
R & 9  & 9  & 0 & 0 & --- \\
Q & 1  & 1  & 0 & 0 & 0\% \\
\bottomrule
\end{tabular}
\end{table}

Removing the two non-member objects and the nine non-comparable R objects leaves 50 names testable on the D/Q axis; exactly one changes class --- 2005 TH173, Q$\to$D. The single D$\to$Q entry in Table~\ref{tab:census} is SN55, already excluded on membership grounds. Splitting the 50 by 2007 orbit quality, none of the 30 multi-opposition objects changes class, versus one of 20 short-arc objects (Fisher's exact two-sided $p=0.40$ on that $1/20$ versus $0/30$ table). Every directional change in the 61-object census --- one reclassification and two population departures --- falls among objects BM09 classified from short arcs; treating those three events as $3/22$ versus $0/30$ gives $p=0.07$. Neither test is statistically significant.

\subsection{Orbit-revision mechanism}
\label{sec:od}

\begin{figure*}[tbp]
\centering
\includegraphics[width=0.72\textwidth]{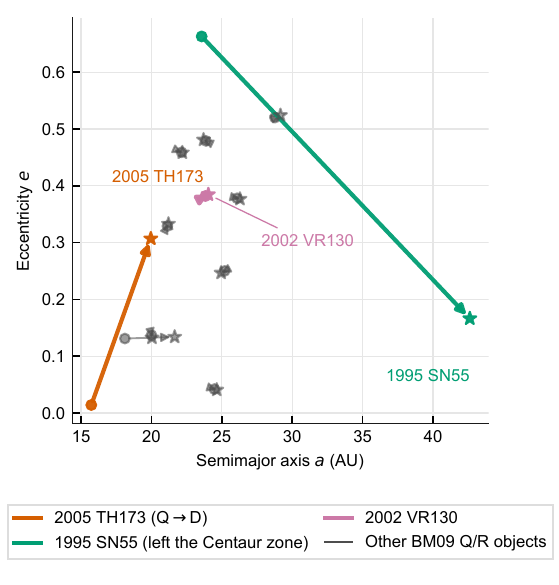}
\caption{BM09 Table~2 $\to$ modern SBDB $(a,e)$ revision arrows for the ten Q/R objects with SBDB matches, plus SN55 and VR130 from the D census (tails: BM09 2007; heads: modern SBDB, osculating epoch 2026-06-09 for most objects, 2003-09-24 for QP112). SN55 has the largest $|\Delta e|$ and $|\Delta a|$ in the sample; TH173 has the largest signed $\Delta e$. VR130 is shown only as a small D-census revision; the production census keeps it as D.\label{fig:od}}
\end{figure*}

The clearest class changes track the largest element revisions (Table~\ref{tab:od}); SN55 has the largest $|\Delta e|$ ($0.497$) and $|\Delta a|$ ($19.049$\,AU), while TH173 has the largest signed $\Delta e$ ($+0.293$). $\Delta a$ and $\Delta e$ are differences of osculating elements at different epochs (BM09 2007-03-06; modern 2026-06-09, or 2003-09-24 for QP112), so part of each includes osculating drift. Figure~\ref{fig:od} depicts these revisions as $(a,e)$ arrows.

\begin{table*}[t]
\centering
\small
\caption{Lead orbit-revision cases on the production census and the 20-clone Q/R matrix. $\Delta a$ and $\Delta e$ are modern minus BM09, as differences of osculating elements at different epochs (BM09 2007-03-06; modern 2026-06-09, or 2003-09-24 for QP112), so part of each includes osculating drift. RO43 is a 10/10 tie in the 20-clone matrix.\label{tab:od}}
\begin{tabular}{@{}lccccc@{}}
\toprule
Object & BM09$\to$v2 & $\Delta a$ (AU) & $\Delta e$ & BM09 arc & Modern (d) \\
\midrule
2005 TH173 & Q rejected & $+4.228$ & $+0.293$ & 17\,d & 366 \\
1995 SN55 & D$\to$Q (not a Centaur) & $+19.049$ & $-0.497$ & 36\,d & 7708 \\
2003 UW292 & R$\to$D & $+3.538$ & $+0.003$ & 28\,d & 8146 \\
2005 RO43 & R$\to$tie & $+0.391$ & $+0.004$ & 3 opp & 7854 \\
2003 QP112 & R$\to$D & $+0.077$ & $+0.004$ & 54\,d & 54 \\
\bottomrule
\end{tabular}
\end{table*}

2002 VR130 (BM09 class D) is not in Table~\ref{tab:od}. The 10-clone production census keeps it as D (8\,D~/~2\,R).

QP112's modern arc equals its BM09 arc (54\,d), so its R/Q minority cannot be attributed to arc growth alone; its SBDB $1\sigma$ uncertainties remain large, so diagonal clones span a wide strip of Centaur $(a,e)$. We report the modal class (D) alongside the full clone counts rather than collapsing QP112 to a single label.

Orbit revision accounts for TH173 and SN55 but is not a clean rank-order predictor of R-minority size across the ten Q/R objects.

\subsection{Orbit update versus pipeline choice}
\label{sec:expc}

A control experiment holds the integrator, escape criteria, angles, and classifier fixed, testing whether substituting modern $(a,e,i)$ for BM09 Table~2 $(a,e,i)$ recovers BM09's published R/Q labels; each branch is a single nominal trajectory at $T_{\max}=40$\,Myr. Under v2, the BM09-element branch recovers 1/10 published R/Q labels and the modern-element branch 3/10 --- updated elements move single-nominal labels toward BM09's classes, the opposite direction from the 20-clone ensembles (Table~\ref{tab:qr}). The suppression of R in those ensembles is attributable to the lifetime gate in the classification rule, not to orbit revision; single-nominal control results do not transfer to ensemble-level classes.

1995 SN55 confirms the production rule can emit Q given genuinely long-lived, flat $a(t)$ behaviour; it is not a Centaur lifetime measurement. Resonant trans-Neptunian objects are expected to be long-lived \citep{1995AJ....110.3073D,2008ssbn.book...43G}, so the Q label here reflects that expectation rather than a rediscovery of the Centaur Q niche.

\section{Discussion}

For objects with arc $\ge 30$\,d, further orbital precision is the wrong lever for lifetime-based classification. The measured leftover dispersion is 0.54\,dex among 65 complete-case Centaurs; conditioned on osculating elements, the slope on $\sigma_a/a$ is consistent with zero, with an effect of up to 0.24\,dex across the seven decades of $\sigma_a/a$ still permitted at 95\% confidence. Follow-up astrometric effort is better directed toward the 55 catalogued Centaurs with arc $<30$\,d --- obtaining first orbits --- than toward refining classes for the 355 objects already past that threshold.

This split follows directly from the two research questions. Lifetime dispersion among well-observed objects is chaos-limited, and BM09's D/Q labels for well-observed objects are correspondingly stable --- none of the 30 multi-opposition 2007 orbits changes class. The labels that do move are short-arc 2007 entries, where the orbit rather than the Lyapunov time was the limiting factor.

The D/Q comparison retains 49/50 remaining Centaurs; TH173 is the exception because its modern solution is Uranus-crossing, with clones escaping on a few-Myr median. BM09's qualitative map of diffusion, sticking, and quasi-stability remains a useful framework \citep{2009Icar..203..155B,2007Icar..190..224D,2003AJ....126.3122T}, but applying it to a named object depends on the orbit of the day and on whether R denotes ``resonance hopping'' or ``survived 22\,Myr.'' The nine BM09 R objects fall outside this test, and these ensembles were not designed to estimate Jupiter-family-comet injection rates or impact probabilities \citep{2019ApJ...883L..25S,2019MNRAS.482..771G}.

\section{Limitations}
\label{sec:limits}

\begin{itemize}
\item Production Q/R tables resample independent Gaussian shifts on $1\sigma$ element uncertainties, not the full SBDB covariance; a 20-clone TH173 covariance check at 10\,Myr still returns 0\,Q, but full-covariance clones could still shift minority R/Q fractions for short-arc objects such as QP112.
\item Orbital elements, integrator, escape criteria, and classifier all differ from BM09; a modern label change is not attributable to astrometry alone.
\item A 10\,Myr stop can confirm early escape but cannot assign R or Q; the 40\,Myr TH173 run (88\,D / 12\,R / 0\,Q) is the threshold test.
\item The production rule was adopted after a Hurst-gated stop-check rather than pre-registered; TH173's Q rejection survives under any lifetime-faithful rule.
\item The BM09 Table~2 census uses ten clones per object; the 20-clone Q/R matrix remains the deeper sample for minority fractions.
\item Classifier v2 assigns R only above the 22\,Myr cut, whereas BM09's R is a morphological statement about resonance hopping; the nine BM09 R objects are excluded from the like-for-like comparison, and a lifetime-independent resonance-angle diagnostic was not implemented.
\item On the D/Q-comparable split the $2\times 2$ table is one class change among 20 short-arc objects versus none among 30 multi-opposition objects (Fisher's exact two-sided $p=0.40$). Counting the two membership losses as additional short-arc events gives $3/22$ versus $0/30$ ($p=0.07$). Neither is significant; the $0/30$ multi-opposition control is the more robust half of this comparison.
\item The four-giant model follows classical Centaur lifetime experiments; close encounters with Earth, Venus, and Mars are not resolved.
\item The Lyapunov proxy does not certify $\tau_{\mathrm{proxy}}$ as a variational Lyapunov time.
\end{itemize}

\section{Conclusions}

Tighter published orbital precision does not reduce the dispersion of simulated Centaur lifetimes once an object has arc $\ge 30$\,d. Among 65 such Centaurs with no censored clones, this dispersion is 0.54\,dex in $\log_{10}$ lifetime --- a factor of 3.5. Conditioned on osculating $a$, $e$, $i$, and $q$, the slope on $\sigma_a/a$ is $-0.002$\,dex per decade in uncertainty (95\% CI $-0.034$ to $+0.030$), which we interpret as a chaos floor; the interval permits a small remaining effect but does not establish exact independence. Of 410 catalogued Centaurs with $q>5.2$\,AU, 355 already have arc $\ge 30$\,d, and follow-up astrometry is most valuable for the remaining 55.

On modern SBDB orbits, 49/50 objects that remain Centaurs and are comparable with BM09 on the D/Q axis retain their class. 2005 TH173 is the exception: 80/100 clones escape before 10\,Myr (median 2.99\,Myr), and a 40\,Myr integration of the same initial conditions yields 88\,D / 12\,R / 0\,Q. None of the 30 multi-opposition 2007 orbits changes class; two short-arc 2007 objects leave the Centaur zone; and 2005 RO43 is a 10/10 D/R tie, not a D majority.

The largest open issue is morphological R: our rule cannot test BM09's resonance-hopping class, and a lifetime-gated letter R should not be equated with it. Full-covariance clones for the short-arc Q/R objects remain a second open issue.

\backmatter

\section*{Statements and Declarations}

\paragraph{Funding.}
This research did not receive any specific grant from funding agencies in the public, commercial, or not-for-profit sectors.

\paragraph{Ethics approval.}
Not applicable.

\paragraph{Consent for publication.}
Not applicable.

\paragraph{Competing interests.}
The authors declare that they have no competing interests.

\paragraph{Author contributions.}
Naresh Prasanna: conceptualization, methodology, software, formal analysis, investigation, data curation, writing --- original draft, visualization. Chrisphin Karthick: conceptualization, supervision, writing --- review and editing.

\paragraph{Materials availability.}
Not applicable.

\paragraph{Code availability.}
Python integration and classification code accompany this submission as archived production scripts, permanently archived on Zenodo \citep{2026zenodo.22848047} (version 1.0.1, DOI: \url{https://doi.org/10.5281/zenodo.22848047}; the concept DOI \url{https://doi.org/10.5281/zenodo.22840759} always resolves to the latest version).

\paragraph{Data availability.}
All production integrations use random seed 87 and SBDB elements retrieved 2026-07-26 (SBDB osculating epoch 2026-06-09, JD 2461200.5, for most objects; 2003-09-24, JD 2452906.5, for 2003 QP112). The archived production tree includes Python integration and classification code, machine-readable BM09 Table~2 and cached modern SBDB element files, and JSON matrices for TH173 (100-clone 10\,Myr and 100-clone 40\,Myr), the Q/R full-scale sample (20-clone, 40\,Myr), the BM09 Table~2 census with 40\,Myr extensions for right-censored clones, the TH173 SBDB-covariance Q-test (20-clone, 10\,Myr), the orbit-update control, the 219-ensemble dispersion analysis, the Lyapunov proxy, the Appendix~\ref{app:e} synthetic-arc-length scan (7 objects $\times$ 5 tiers $\times$ 20 clones), and the Appendix~\ref{app:f} population survey. The production archive (title: Centaur Longevity Revisited: Lifetime Dispersion and the Bailey \& Malhotra Classifications under Modern Orbital Solutions --- Reproducibility Archive) is permanently archived in the Zenodo repository at \url{https://doi.org/10.5281/zenodo.22848047} \citep{2026zenodo.22848047} (version 1.0.1). The concept DOI \url{https://doi.org/10.5281/zenodo.22840759} always resolves to the latest version. The archive is also available at \url{https://github.com/naresh-prasanna/Centaur_Longevity_Revisited}.

\paragraph{Use of generative AI.}
During manuscript preparation the authors used Cursor (Anysphere; Composer~2.5 and Grok~4.6) to draft and revise text, check numbers against archived data products, and format tables and references. After that use the authors reviewed and edited the content and take full responsibility for it. These tools are not authors. Research-process use of the same tool is described in \S\ref{sec:software}.

\begin{appendices}
\renewcommand{\thefigure}{\arabic{figure}}
\renewcommand{\thetable}{\arabic{table}}
\setcounter{figure}{8}
\setcounter{table}{10}

\section{Hurst-gated classifier (v1) as a robustness check}
\label{app:v1}

Production tables use the lifetime-first rule (v2). An earlier Hurst-gated tree assigned D whenever the linear Hurst fit was sufficiently good ($R^{2}\ge 0.85$), even for clones already surviving 22\,Myr. Relabelling the same archived 20-clone, 40\,Myr lifetimes under v2, without re-integration, yields D as the modal class for nine of ten Q/R objects; RO43 remains a 10/10 D/R tie (Table~\ref{tab:v1v2}). FZ53 and QP112 stay D (16/4/0 and 11/7/2), matching Table~\ref{tab:qr}. The production rule was adopted because it follows BM09's stated lifetime-first ordering; this choice was not pre-registered. TH173's Q rejection is unchanged under either rule.

\setcounter{table}{10}
\begin{table}[!ht]
\centering
\small
\caption{Q/R class under the Hurst-gated rule versus the production lifetime-first rule (same 20-clone, 40\,Myr integrations; re-label only).\label{tab:v1v2}}
\begin{tabular}{@{}lccc@{}}
\toprule
Object & BM09 & Hurst-gated & Production \\
\midrule
1995 DW2 & R & D & D \\
1998 QM107 & R & D & D \\
1998 TF35 & R & D & D \\
2000 FZ53 & R & D & D \\
2003 QP112 & R & D & D \\
2003 UW292 & R & D & D \\
2005 RL43 & R & D & D \\
2005 RO43 & R & D & tie \\
2005 TH173 & Q & D & D \\
2006 SX368 & R & D & D \\
\midrule
Agree with BM09 & --- & 0/10 & 0/10 ($+$1 tie) \\
\bottomrule
\end{tabular}
\end{table}
\FloatBarrier

\section{Synthetic arc length and class mixture}
\label{app:e}

This appendix tests whether synthetic arc length --- via $\sigma \propto \sqrt{A/\mathrm{arc}}$ at fixed modern nominal elements --- alters class mixture. Design: 7 objects $\times$ 5 arc tiers $\times$ 20 clones, $T=25$\,Myr, classifier v2, seed 87. The seven objects are 2005 TH173, 1995 DW2, 1999 HD12, 1996 RX33, 2003 QP112, 1998 QM107, and 1977 UB. The scan varies formal uncertainty at fixed nominal elements only; real arc growth also alters the nominal solution and covariances.

D remains the modal class at every arc tier for all seven objects; where trends appear, they affect minority fractions rather than the mode (Fig.~\ref{fig:arc}).

For 1999 HD12 (Fig.~\ref{fig:arc}b), the same five-tier scan returns D/R/Q $=12/0/8$, $12/0/8$, $12/0/8$, $13/0/7$, and $14/0/6$ at 10, 15, 22, 33, and 49\,d; D remains modal while the Q minority falls.

TH173 synthetic tiers (20 clones, $T=25$\,Myr): at 30, 56, 105, 196, and 366\,d, D/R/Q $=19/1/0$, $19/1/0$, $19/1/0$, $18/2/0$, and $17/3/0$ respectively; Spearman(arc, minority R) $\approx +0.89$ across the five tiers, with D remaining modal throughout.

\setcounter{figure}{8}
\begin{figure*}[t]
\centering
\includegraphics[width=0.88\textwidth]{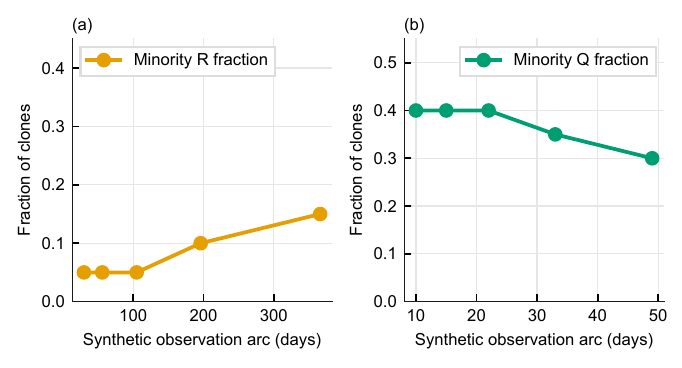}
\caption{Exploratory arc-to-mixture scan (20 clones/tier, $T=25$\,Myr, classifier v2, seed 87). (a)~2005 TH173: minority-R fraction rises across five synthetic arc tiers while D remains modal. (b)~1999 HD12: minority-Q fraction falls under the same design while D remains modal.\label{fig:arc}}
\end{figure*}

\section{A null survey of orbit quality against classification fragility}
\label{app:f}

Preceding the dispersion analysis of \S\ref{sec:chaosfloor}, we also tested whether orbit-determination quality predicts class-based fragility across the Centaur population; this design failed for want of variance in the outcome. From the JPL SBDB Centaur catalogue we drew 172 objects stratified by arc length, condition code, and estimated opposition count, each receiving ten diagonal clones to $T_{\max}=25$\,Myr. All 172 objects return D as the modal class, since almost no clone reaches the 22\,Myr cut within that window. The substantive bound on orbit quality is therefore the leftover-dispersion analysis in the main text, not this class-based survey.

\end{appendices}

\end{document}